\documentstyle[galley,epsf]{mn}
\newif\ifAMStwofonts

\ifoldfss
  \ifCUPmtlplainloaded \else
    \NewTextAlphabet{textbfit} {cmbxti10} {}
    \NewTextAlphabet{textbfss} {cmssbx10} {}
    \NewMathAlphabet{mathbfit} {cmbxti10} {} 
    \NewMathAlphabet{mathbfss} {cmssbx10} {} 
  \fi
  \ifAMStwofonts
    \ifCUPmtlplainloaded \else
      \NewSymbolFont{upmath} {eurm10}
      \NewSymbolFont{AMSa} {msam10}
      \NewMathSymbol{\upi}     {0}{upmath}{19}
      \NewMathSymbol{\umu}     {0}{upmath}{16}
      \NewMathSymbol{\upartial}{0}{upmath}{40}
      \NewMathSymbol{\leqslant}{3}{AMSa}{36}
      \NewMathSymbol{\geqslant}{3}{AMSa}{3E}

    \fi
  \fi
\fi 

\ifnfssone
  \newmathalphabet{\mathit}
  \addtoversion{normal}{\mathit}{cmr}{m}{it}
  \addtoversion{bold}{\mathit}{cmr}{bx}{it}
  \newmathalphabet{\mathbfit} 
  \addtoversion{normal}{\mathbfit}{cmr}{bx}{it}
  \addtoversion{bold}{\mathbfit}{cmr}{bx}{it}
  \newmathalphabet{\mathbfss} 
  \addtoversion{normal}{\mathbfss}{cmss}{bx}{n}
  \addtoversion{bold}{\mathbfss}{cmss}{bx}{n}
  \ifAMStwofonts
    \ifCUPmtlplainloaded \else
      \UseAMStwoboldmath
      \makeatletter
      \new@mathgroup\upmath@group
      \define@mathgroup\mv@normal\upmath@group{eur}{m}{n}
      \define@mathgroup\mv@bold\upmath@group{eur}{b}{n}
      \edef\UPM{\hexnumber\upmath@group}
      \new@mathgroup\amsa@group
      \define@mathgroup\mv@normal\amsa@group{msa}{m}{n}
      \define@mathgroup\mv@bold\amsa@group{msa}{m}{n}
      \edef\AMSa{\hexnumber\amsa@group}
      \makeatother
      \mathchardef\upi="0\UPM19
      \mathchardef\umu="0\UPM16
      \mathchardef\upartial="0\UPM40
      \mathchardef\leqslant="3\AMSa36
      \mathchardef\geqslant="3\AMSa3E
    \fi
  \fi
\fi 

\ifnfsstwo
  \DeclareMathAlphabet{\mathbfit}{OT1}{cmr}{bx}{it}
  \SetMathAlphabet\mathbfit{bold}{OT1}{cmr}{bx}{it}
  \DeclareMathAlphabet{\mathbfss}{OT1}{cmss}{bx}{n}
  \SetMathAlphabet\mathbfss{bold}{OT1}{cmss}{bx}{n}
  \ifAMStwofonts
    \ifCUPmtlplainloaded \else
      \DeclareSymbolFont{UPM}{U}{eur}{m}{n}
      \SetSymbolFont{UPM}{bold}{U}{eur}{b}{n}
      \DeclareSymbolFont{AMSa}{U}{msa}{m}{n}
      \DeclareMathSymbol{\upi}{0}{UPM}{"19}
      \DeclareMathSymbol{\umu}{0}{UPM}{"16}
      \DeclareMathSymbol{\upartial}{0}{UPM}{"40}
      \DeclareMathSymbol{\leqslant}{3}{AMSa}{"36}
      \DeclareMathSymbol{\geqslant}{3}{AMSa}{"3E}
    \fi
  \fi
\fi 

\ifCUPmtlplainloaded \else
  \ifAMStwofonts \else 
    \def\upi{\pi}
    \def\umu{\mu}
    \def\upartial{\partial}
  \fi
\fi

\begin{document}
\title{High-resolution spectroscopy of QY Sge -- An obscured RV Tauri variable?}
\author[N. Kameswara Rao, Aruna Goswami,  \& David L. Lambert]
       {N. Kameswara Rao,$^1$ Aruna Goswami,$^1$ David L. Lambert,$^2$\\ 
       $^1$Indian Institute of Astrophysics, Bangalore 560034, India\\
       $^2$Department of Astronomy, University of Texas, Austin, TX 78712-1083, USA\\}
\date{Accepted .
      Received ;
      in original form 2001 }

\pagerange{\pageref{firstpage}--\pageref{lastpage}}
\pubyear{2001}

\maketitle

\label{firstpage}

\begin{abstract}

The first  high-resolution optical spectra of QY Sge are  presented and
discussed.  Menzies \& Whitelock (1988) on the basis of  photometry and
low-resolution spectra suggested that this G0I supergiant  was obscured by dust
and seen only by scattered light from a
circumstellar reflection nebula. The
new spectra confirm and extend this picture.
Photospheric lines are unusually broad indicating scattering of photons
from dust in the stellar wind. Presence of very  broad Na D emission 
lines   is confirmed. Sharp emission lines from low levels
of abundant neutral metal atoms are reported for the
first time.
An abundance analysis of photospheric lines
 shows that the stellar atmosphere is of approximately solar 
composition but with highly condensible (e.g., Sc and Ti) elements depleted  
by factors of 5 to 10.

\end{abstract}


\section{Introduction}

Our  investigation of QY\,Sge (IRAS\,20056+1834) was
prompted by Menzies \& Whitelock's (1988) report of
broad Na\,D emission lines in the spectrum of this 
G0 supergiant with a strong   infrared excess.
 Broad Na D (and other) emission lines
are seen in  spectra of R\,Coronae Borealis stars taken 
at minimum light. Questions abound about the location of the broad-line
emitting gas with respect to the R\,CrB star, and about the
excitation mechanism for these lines.
Insights into the  formation of the broad lines
may possibly be obtainable by observing other
stars exhibiting similar lines, particularly,  targets like
QY Sge where the emission lines seem to be a permanent feature of the
spectrum. It was in this spirit that we observed 
QY\,Sge at high spectral
resolution.

 Menzies \&
Whitelock observed QY Sge  at 3.5\AA\ resolution  at several epochs in the
late 1980s. The Na\,D lines were seen strongly in emission with a width
of 140 km s$^{-1}$ (FWHM) and at a  velocity  equal to
the stellar absorption line velocity to within the
errors of measurement ($\pm$ 15 km s$^{-1}$). Other evidence for
emission was offered: Ca\,{\sc ii} H and K showed emission
cores, Ca\,{\sc i} 4226\AA\ was weakly in emission,
and H$\alpha$ was filled in by emission. Consideration of the
absorption lines led to a spectral type of
about G0I. 

QY Sge's infrared excess signals the presence of a thick dust cloud  for
which Menzies \& Whitelock estimated a  blackbody
temperature of about 600 K.
 Polarimetry at optical (Trammell,
Dinerstein, \& Goodrich 1994) and infrared (Gledhill et al. 2001)
wavelengths showed significant polarization with the degree of
polarization increasing from the visual to a maximum of about 14 \% in the J
 band and then 
decreasing to about 7 \% in the H band and  3 \% in the K band.
Optical polarimetry provides an important clue to the location of the
gas emitting the broad lines;
the broad Na\,D emission  lines are unpolarized although the local continuum
is polarized (Trammell et al. 1994).
 At about the H band and to longer
wavelengths, the flux is principally emission from the dust, but
at wavelengths shorter than H band, the flux is photospheric
radiation scattered off the surrounding dust (Menzies \& Whitelock 1988).
The source has not been spatially resolved.

The parallels with the R\,CrB stars are several -- yellow supergiant,
dusty circumstellar cloud, and broad emission lines -- but 
incomplete in that QY\,Sge is not obviously H-deficient.
 The R\,CrB stars' broad emission lines
are seen only after a star has faded by several magnitudes. Yet,
QY\,Sge shows the broad lines at `maximum' light.
This happy circumstance arises because dust dims the star but not
the region emitting the broad lines.
While there may be similarities of geometry and excitation
between QY Sge and the R CrB stars, their evolution is likely to
have differed greatly. Menzies \& Whitelock (1988) discussed several
possible histories for QY Sge but were unable to decide whether
the star was an evolved massive star or a low mass post-AGB star.
Others (Trammell et al. 1994; Gledhill et al. 2001) have
 assumed that QY Sge should be listed with   post-AGB stars and
protoplanetary nebulae. 

\section{Observations}

QY\,Sge was observed with the McDonald Observatory's
2.7m Harlan J. Smith telescope and its `2dcoud\'{e}'
cross-dispersed echelle spectrograph
(Tull et al. 1995) on  1999 August 17 and 18, and
2000 June 14. The  observed bandpass ran from about 3900 \AA\ to
10000 \AA\ with gaps beyond about 5600 \AA\ where the echelle
orders were incompletely captured on the Tektronix 2048 $\times$ 2048
CCD.
 An additional observation was obtained on 2001 July
13 using the Sandiford Cassegrain echelle spectrograph on the McDonald
2.1m Otto Struve 
 reflector (McCarthy et al. 1993). This spectrum recorded with
a Reticon CCD provided complete
coverage of the
interval 4500 \AA\ to 5170 \AA.
  Observations of a Th-Ar hollow cathode lamp provided the
wavelength calibration. All spectra were obtained at a nominal spectral
resolving power of 60 000.
Data were reduced in the standard fashion
using the IRAF\footnote{IRAF is distributed by the National Optical
Astronomical Observatories, which is operated by the Association
for Universities for Research in Astronomy, Inc., under contract
to the National Science Foundation.} software package.

Spectra of
$\gamma$\,Cyg, and 89\,Her were obtained also with the `2dcoud\'{e}'
spectrograph.
Gamma Cyg is a normal F8Iab star.
The spectrum of 89\,Her,  a
warmer supergiant (spectral type F2Ib),  includes
sharp emission lines from low excitation levels of 
abundant atoms, but lacks  broad emission lines.

\section{The Absorption and Emission Line Spectrum}

Inspection of QY Sge's spectra shows three obvious components:
 absorption lines representative of
a late-F to early-G  supergiant, sharp emission lines from resonance and low
excitation transitions of abundant neutral atoms, 
and a few broad emission 
lines (e.g., Na\,D and K\,{\sc i} 7665 \AA\
and 7699 \AA\ lines).

\subsection{The Absorption Lines}

\begin{figure}
\epsfxsize=8truecm
\epsffile{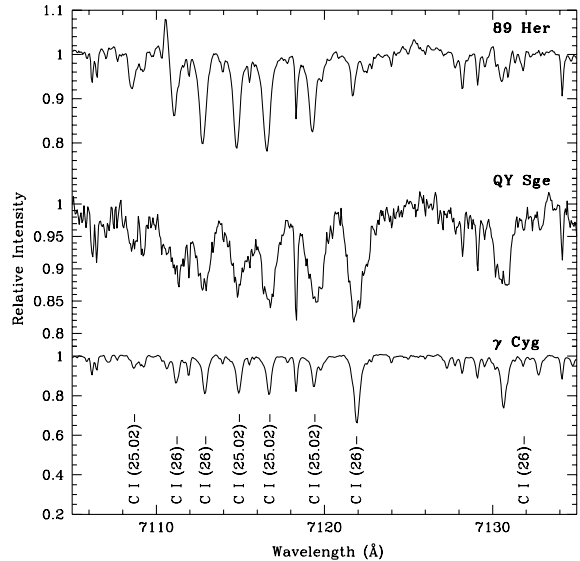}
\caption{Spectra of 89 Her, QY Sge, and $\gamma$ Cyg from 7105 \AA\ to 7135 \AA.
Lines of neutral carbon are  prominent
in this interval. Equivalent widths of C\,{\sc i} lines are considerably
larger in QY Sge than in $\Gamma$ Cyg.
}
\end{figure}

The absorption line spectrum broadly resembles that of $\gamma$ Cyg.
A striking difference is that  QY Sge's lines
are much broader.
A few lines are clearly stronger in QY Sge than in $\gamma$ Cyg.
 Among these
 are the C\,{\sc i} lines 
(Figure 1). 
Low excitation lines of neutral metals are also
stronger in QY Sge.
 A spectacular example is shown in Figure 2:
the Fe\,{\sc i} at 5060.1 \AA\ from the resonance multiplet. 
 This and similar lines are slightly
red-shifted relative
to higher excitation lines (see below). 
A few lines  are
greatly weaker in QY Sge. These are lines of singly-ionized atoms
such as Sc\,{\sc ii}, Ti\,{\sc ii}, Y\,{\sc ii}, and
Nd\,{\sc ii}. Three examples are given in Figure 2. 
Line weakening does not appear to be due to overlying emission.

\begin{figure}
\epsfxsize=8truecm
\epsffile{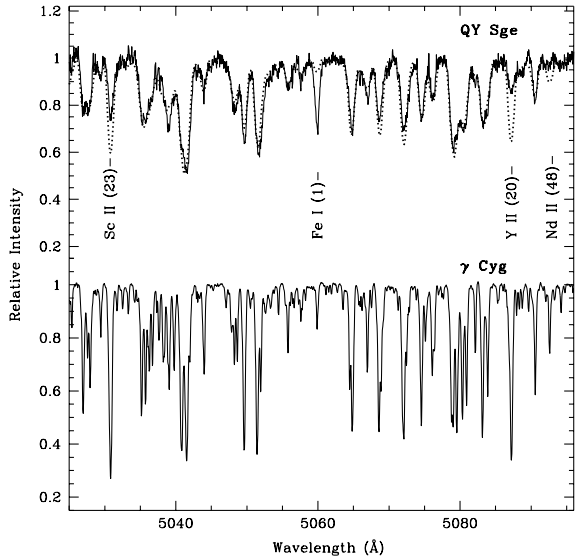}
\caption{The spectrum of QY Sge (upper panel) from 5025 \AA\ to 5095 \AA\
with the spectrum of $\gamma$ Cyg (lower panel). The dotted line in the
upper panel is the spectrum of $\gamma$ Cyg broadened by 30 km s$^{-1}$
(see text). Many lines in the broadened spectrum match well the
strengths of the same lines in QY Sge. Notable exceptions are the
Fe\,{\sc i} RMT1 resonance line, which is much stronger in QY Sge, and
lines of Sc\,{\sc ii}, Y\,{\sc ii}, and Nd\,{\sc ii},
which are weaker in QY Sge.}
\end{figure}


By inspection, $\gamma$\,Cyg's lines are sharper than those in
QY Sge.
To facilitate a direct comparison of  the spectra,
we have broadened $\gamma$\,Cyg's spectrum to match
that of QY\,Sge. Assuming that a standard profile for
a rigidly rotating star suffices, we find that a projected velocity
$v\sin i = 30 \pm$ 5 km s$^{-1}$ equalizes the line widths in the
two spectra.
This exercise should not be taken to imply that rotational
line
broadening is necessarily responsible for QY Sge's broader lines.
The difference cannot  be attributed to a higher microturbulence
in QY Sge's atmosphere (see below).
Alternatively, if  macroturbulence is invoked, the motions in
QY Sge are highly supersonic, an improbable scenario.  
As we discuss below, the broadening is plausibly attributed to the 
receipt of photons not directly from the star but  after
scattering off moving dust grains in the stellar wind.


The radial velocity was measured from a set of unblended absorption
lines. For the `2dcoud\'{e}' spectra about 150 lines were measured.
Fewer lines - about 30 - were selected from the Sandiford spectrum.
The radial velocities are -21 $\pm$ 1.5 km s$^{-1}$ for 1999 August 17
and 18, -23 $\pm$ 1.5 km s$^{-1}$ for 2000 June 14,
and -9.1 $\pm$ -2.2 km s$^{-1}$ for 2001 July 13.
 The star is obviously
a velocity variable which is not surprising for such a luminous star.
At this time, the systemic velocity is unknown.

Low excitation Fe\,{\sc i} and other
 lines show a more positive velocity by about
 7 to 9  km s$^{-1}$.
While the equivalent widths of the
 higher excitation lines agree well with those of the lines in
$\gamma$ Cyg, the low excitation lines with the more positive
velocity are stronger in QY Sge than
in $\gamma$ Cyg. This difference rules out  emission line
contamination of QY Sge's low excitation lines.
These measurements refer to spectra taken in  1999 and 2000.

\begin{figure}
\epsfxsize=8truecm
\epsffile{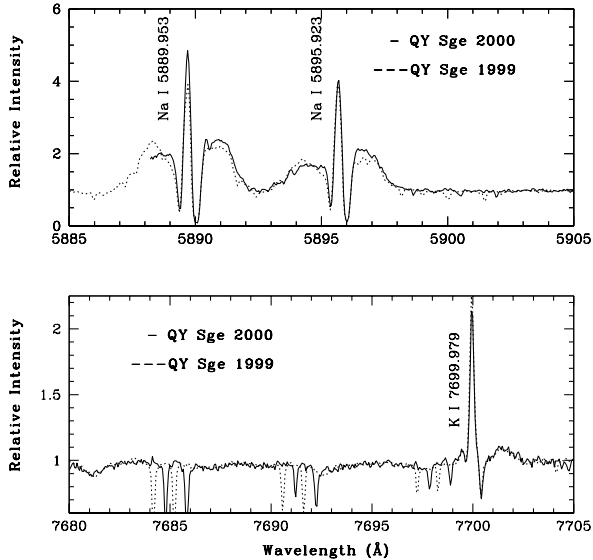}
\caption{Spectra of QY Sge in 1999 August and 2000 June showing the
emission profiles of the Na\,D lines (upper panel) and the K\,{\sc i}
resonance lines (lower panel). Note the presence of sharp and broad
components for these resonance lines. The sharp absorption lines, especially
prominent in the lower panel, are of telluric origin.}
\end{figure}

\begin{figure}
\epsfxsize=8truecm
\epsffile{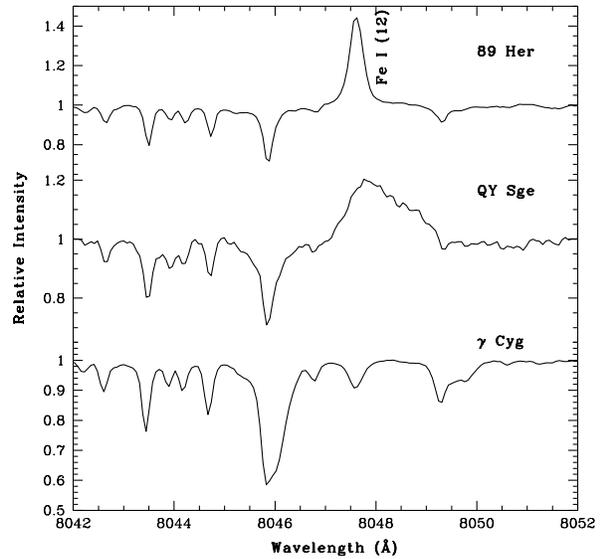}
\caption{ Spectra of 89 Her, QY Sge, and $\gamma$ Cyg from 7970 \AA\ to 8060
 \AA. The prominent emission line of Fe\,{\sc i} from RMT12 is
seen in QY Sge and 89 Her. }
\end{figure}

\subsection{Sharp Emission Lines}
 
The most prominent of the sharp emission lines are the Na\,D and
K\,{\sc i} resonance lines (Figure  3). Weaker lines 
are scattered across the bandpass. These low excitation
lines are less common than in the spectrum of 89 Her, and
also broader than their 89 Her counterparts. Among 
the  weaker lines are the Fe\,{\sc i} lines of
multiplet RMT(12) of which an example is shown in Figure 4.

The radial velocity  of the emission peak for the strongest
lines is 
-21  km s$^{-1}$ from the Na\,D lines, -24  km s$^{-1}$
from the K\,{\sc i} lines, and -24 km s$^{-1}$ from the Ca\,{\sc ii} H line.
These results are from the 1999 August spectrum but the 2000 June
spectrum gives similar results. The peak velocity is essentially
equal to
the absorption line
velocity given above. This may be a coincidence because it
appears that these strong emission lines are flanked by
absorption, a suspicion reinforced by the fact that the
widths (7 km s$^{-1}$) of the emission lines are much smaller than
those (26 km s$^{-1}$) of weaker lines.

The weak
 emission lines of Fe\,{\sc i}  - RMT(12) and RMT(13) - are at a velocity
of about -14 km s$^{-1}$, which is the velocity of the
absorption lines from the same low excitation levels. and
  red-shifted  by about
7 km s$^{-1}$ with respect to the
absorption line velocity measured from higher excitation lines.  
These emission lines are broad (FWHM $\simeq$ 26 km s$^{-1}$) relative
to their counterparts in 89 Her.
The upper levels of the absorption lines of Fe\,{\sc i} RMT(1) serve as the
upper level of the emission lines from Fe\,{\sc i} RMT(12). Similarly,
the upper level for Fe\,{\sc i} RMT(2) seen in absorption feeds Fe\,{\sc i}
RMT(13) seen in emission;  iron atoms from low lying states are
absorbing photons and feeding them into other lines sharing the
upper state. All of these measurements refer to the 1999 and 2000
spectra.

\subsection{Broad Emission Lines}

What drew us to study QY Sge was the report of  broad emission lines. 
The most prominent are the Na\,D  lines which include a sharp line (Figure
3). This Figure also shows the K\,{\sc i} resonance
lines where the broad component is present but less prominent
relative to the sharp emission component than
in the Na D lines.
Broad emission is weakly present in the Ca\,{\sc ii} H line.
Figure 5 compares the Na D profiles of QY Sge with profiles for three
other stars: R CrB at
minimum light, 89 Her, and the normal supergiant $\gamma$ Cyg.
Gamma Cyg shows the photospheric Na D profile with two weak
blue-shifted absorption lines in the line cores; these are
possibly interstellar components. For 89 Her, the photospheric
Na D absorption lines appear partly filled in with emission, and
complex blue-shifted absorption  present at about 2 \AA\ from the
deep absorption core is probably of circumstellar origin. 
Very weak emission may be
present in the red wing of the D2 line.
Broad emission in addition to sharp emission is seen in QY Sge and
R\,CrB.
For R\,CrB the broad emission of the  red wing of D2 overlaps
the blue wing of D1, but 
D1 and D2 are well resolved  for QY\,Sge. The intrinsic profile of 
the broad lines in QY Sge is probably that of a single peak,
but the superposition of the sharp emission and accompanying
absorption creates the appearance of a broad emission split into
a blue and red component of approximately equal intensity.
A part of the central absorption may be contributed by the
photospheric line. The D1/D2 flux ratio is slightly less than
the value of two expected for optically thin lines.

QY Sge's  broad emissions are approximately centred on the stellar velocity.
Average velocity of the blue and red peaks is -24 km s$^{-1}$ for
the Ca\,{\sc ii} K line, and -27 km s$^{-1}$ for D2, and -29 km s$^{-1}$
for D1 from the 1999 August spectrum for which the systemic velocity
was -21 km s$^{-1}$.
The separation of the blue and red peaks is about 130 km s$^{-1}$.
 The base widths of the lines are about 270 km s$^{-1}$.
Emission in H$\alpha$ may be related to these broad emissions: the P Cygni profile
at H$\alpha$ has red emission extending to -100 km s$^{-1}$ from the
photospheric velocity.

\begin{figure}
\epsfxsize=8truecm
\epsffile{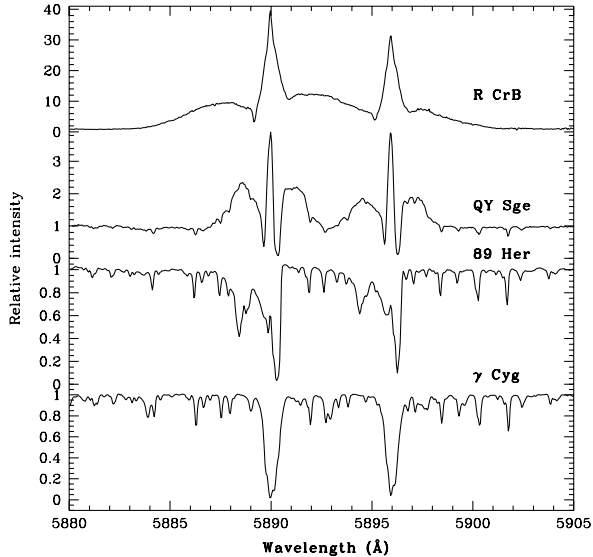}
\caption{The Na D lines in R CrB at minimum light, QY Sge, 89 Her, and
$\gamma$ Cyg. Some of the sharp absorption lines in these spectra are
 telluric H$_2$O lines. }
\end{figure}

\section{An Abundance Analysis}

Although it is recognized  that QY Sge's photosphere may have
an unusual construction,
 it is useful to search
for large abundance anomalies. Quite obviously, the low excitation
lines  and lines contaminated
by overlying emission should be excluded from the abundance
analysis. In particular, we excluded resonance lines and lines
which appear in emission in 89 Her.
A standard LTE model atmosphere-based analysis was undertaken.
Models were taken from Kurucz's grid (http://cfaku5.harvard.edu),
 and the current
version of MOOG (Sneden 1973) was used to predict equivalent widths.

Usual procedures were used to determine the defining
parameters: the effective temperature T$_{\rm eff}$,the surface
gravity  $\log g$, the iron abundance [Fe/H], and
the microturbulence $\xi_t$.
A collection of unblended Fe\,{\sc i} lines was used to
determine T$_{\rm eff}$ and 
 $\xi_t$. Then, a set of Fe\,{\sc ii} lines with the
Fe\,{\sc i} lines and  the assumption of ionization equilibrium gave
$\log g$.
We obtained T$_{\rm eff} =
5850 \pm 200$ K, $\log g = 0.7 \pm 0.25$ in cgs units,
$\xi_t = 4.5 \pm 0.5$ km s$^{-1}$, and [Fe/H] $ = -0.4 \pm 0.1$.
These values are quite similar to those found for $\gamma$ Cyg by Luck \&
Lambert (1981): T$_{\rm eff}$ = 6100 K, $\log g = 0.5$, $\xi_t$ = 3.5 km s$^{-1}$,
and [Fe/H] = +0.1, which is not a surprise given the similarity in the
spectra of the two supergiants.

There is no evidence that QY Sge is hydrogen deficient. Paschen lines
appear in QY Sge with  strengths similar to those of the lines
in $\gamma$ Cyg. The H$\alpha$ profile  has
a central absorption strength comparable 
to that of the photospheric line in $\gamma$ Cyg. Furthermore, as noted
below, the C\,{\sc i} and CH lines return the same carbon abundance.

\begin{table}
\centering
\begin{minipage} {140mm}
\caption{Chemical composition of QY Sge}
\begin{tabular}{lrlrrrr} \hline
Element & Z & $\log\epsilon$ & {N}\footnote{Number of lines} & {$\log\epsilon_\odot$}\footnote{Solar system abundance from Grevesse \& Sauval (1998).} & [X/H] & [X/Fe] \\ \hline
C\,{\sc i}                   &  6      & 8.85$\pm$0.26  &    9     & 8.52         & +0.33        & +0.59        \\
N\,{\sc i}                   &  7      & 8.83$\pm$0.15  &    4     & 7.92         & +0.91        & +1.15        \\
O\,{\sc i}                   &  8      & 9.15  &    1     & 8.83         & +0.32        & +0.59      \\
Na\,{\sc i}                  &  11     & 6.80$\pm$0.21  &    4     & 6.33         & +0.47        & +0.73        \\
Mg\,{\sc i}                 &  12     & 7.44$\pm$0.28  &    2     & 7.58         & -0.14        & +0.12  \\
Mg\,{\sc ii}                 &  12     & 7.52         &    1     & 7.58         & -0.06        & +0.24  \\
Al\,{\sc i}                  &  13     & 5.93         &    1     & 6.48         & -0.55        & -0.29   \\
Si\,{\sc i}                 &  14     & 7.56$\pm$0.20  &   10     & 7.55         & +0.01        & +0.27   \\
S\,{\sc i}                   &  16     & 7.47$\pm$0.10  &    3     & 7.26         & +0.21       & +0.47   \\
Ca\,{\sc i}                  &  20     & 5.97$\pm$0.14  &   11     & 6.36         & -0.39        & -0.13   \\
Ca\,{\sc ii}                &  20     & 6.18$\pm$0.06  &    2     & 6.36         & -0.18        & +0.08   \\
Sc\,{\sc ii}                &  21     & 2.47$\pm$0.22  &    5     & 3.13         & -0.66        & -0.40   \\
Ti\,{\sc ii}                &  22     & 3.98$\pm$0.04  &    3     & 4.98         & -1.00        & -0.74   \\
Cr\,{\sc i}                  &  24     & 5.93$\pm$0.25  &    4     & 5.68         & +0.25        & +0.51   \\
Cr\,{\sc ii}                 &  24     & 5.80$\pm$0.20  &    7     & 5.68         & +0.12        & +0.38   \\
Fe\,{\sc i}                  &  26     & 7.26$\pm$0.22  &    76    & 7.50         & -0.24        &  ....      \\
Fe\,{\sc ii}                 &  26     & 7.22$\pm$0.11  &    12    & 7.50         & -0.28        & ....       \\
Ni\,{\sc i}                  &  28     & 6.01$\pm$0.12  &    8     & 6.25         & -0.24        & +0.02    \\
Zn\,{\sc i}                  &  30     & 4.46  &    1     & 4.63         & -0.17        & +0.09     \\
Y\,{\sc ii}                  &  39     & 1.03$\pm$0.12  &    4     & 2.24         & -1.21        & -0.95    \\
Zr\,{\sc ii}                 &  40     & 1.50         &    1     & 2.60         & -1.10        & -0.84    \\
Ba\,{\sc ii}                 &  56     & 2.07         &    1     & 2.17         & -0.10        & +0.16   \\
Ce\,{\sc ii}                 &  58     & 0.61         &    1     & 1.61         & -1.00        & -0.74   \\
Eu\,{\sc ii}                 &  63     & 0.34$\pm$0.08  &    2     & 0.53         & -0.19        & +0.07   \\
\hline
\end{tabular}
\end{minipage}
\end{table}

Derived abundances are given in Table 1 
as $\log \epsilon$(X) on the scale $\log \epsilon$(H) = 12,
[X/H], and [X/Fe].
The  tabulated uncertainties are derived from the line-to-line
scatter; standard error is smaller by $\sqrt{\rm N}$.
Uncertainties arising from the errors in the model atmosphere
parameters are less than about $\pm$ 0.2 dex.
The fact that the absorption lines are  broad  introduces an above
average uncertainty  in measurements of the equivalent widths.

Carbon and nitrogen are overabundant.
The selection of C\,{\sc i} and N\,{\sc i} lines avoids very 
strong lines.
 As a check on our analysis, we
analysed the C\,{\sc i} lines in our spectrum of $\gamma$ Cyg
obtaining the same abundance as Luck \& Lambert (1981). There are CH lines
in the spectrum of QY Sge. Analysis of these lines returned the carbon
abundance in Table 1. The oxygen abundance is best determined from
the [O\,{\sc i}] lines; the 6363 \AA\ line is present and  useful
despite mild contamination from the telluric emission line.
 The star's photosphere is O-rich not C-rich. 
Sodium's overabundance is based on four low excitation
 lines. The Na\,{\sc i} lines
are clearly stronger than in $\gamma$ Cyg. This is also the
case for the  high-excitation S\,{\sc i} lines. 

For the underabundant elements (Sc, Ti, Y, Zr, and Ce),
 the lines 
 are obviously weaker than
their counterparts in $\gamma$ Cyg.
 The selected lines do not appear in
emission in 89 Her where emission lines are more numerous.
It is striking that the heavy elements have unusual relative abundances:
Y, Zr, and Ce represented by weak lines
 are underabundant by about 1 dex but Ba and Eu show
approximately solar abundances. This anomaly we explain
below. Although the Ba abundance is derived from the
weakest of the observed lines, namely the 5853 \AA\ line with an
equivalent width of  288 m\AA, the Ba abundance is quite sensitive
to the assumed microturbulence. The Y, Zr, Ce, and Eu abundances
are insensitive to the
microturbulence.

These abundance anomalies are reminiscent of a pattern exhibited by
the warmer RV\,Tauri variables whose photospheres are depleted in
those elements which are the first to  condense into grains as gas cools.
Giridhar, Lambert, \& Gonzalez (2000), who
discuss abundances derived for a sample of RV Tauri variables, show
that the abundance anomalies are present only in the warmer stars.
QY Sge's T$_{\rm eff}$ and $\log g$ place it among the affected
variables, as Figure 6 shows.
One may suppose that the  report of photometric
variations on a characteristic time-scale of 50 days (Menzies
\& Whitelock 1988) is a sign that
QY Sge is a RV Tauri variable. The infrared excess of QY Sge
is only slightly more extreme than that of the extreme case -
AR Pup - included by Giridhar et al. in their discussion of
infrared excesses: QY Sge has the infrared colours  J-K = 2.9 and
K-L  = 2.6 (Menzies \& Whitelock 1988)  compared to AR Pup's
J-K = 2.6 and K-L = 2.1. Abundance anomalies are more severe for AR Pup
 than for QY Sge but Giridhar et al. note that these anomalies do
not correlate well with the infrared excesses.

\begin{figure}
\epsfxsize=8truecm
\epsffile{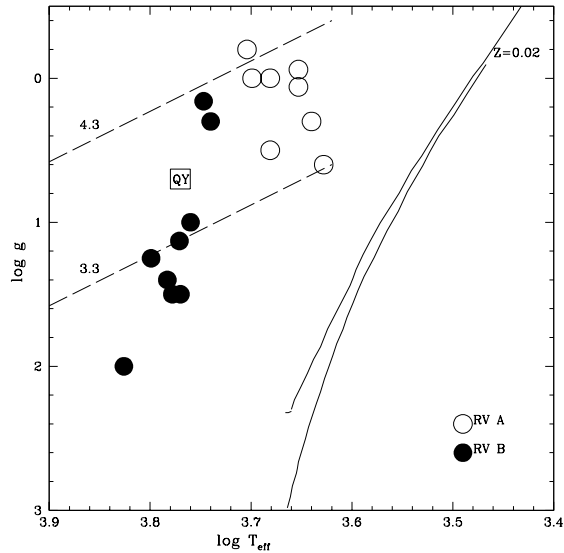}
\caption{The $\log g$ vs. $\log T_{\rm eff}$ diagram for RV Tauri
variables (Preston types RV A and B) and QY Sge. A theoretical isochrone
(solid line) is shown for an age of 10$^{10}$ years and solar
composition ($Z$ = 0.02) from Bertelli et al. (1994):
the red giant branch is represented by the right-most solid line, and the
AGB from the He-core burning stars to the most luminous
AGB stars by the left-most solid line.
 The dashed lines
are tracks for a stellar mass of 0.8 $M_\odot$ evolving at a constant
luminosity of $\log L/L_\odot$ = 3.3 ({\it lower line}) and 4.3 ({\it upper
line}).   }
\end{figure}

Abundance anomalies among the warm RV Tauri variables (Preston types
RV B) are fairly well
correlated with the condensation temperature ($T_{cond}$)
computed for equilibrium
cooling of a solar composition gas at low pressure; the severest underabundances are found for
those elements expected to condense out at the highest temperatures.
 The temperature T$_{\rm cond}$ is that
at which 50 \% of the element has condensed into solids. We take
estimates of T$_{\rm cond}$ from Lodders \& Fegley (1998) and
Wasson (1985). For several elements, the former reference gives the
temperature at which a compound starts forming. In such cases, we have
referred to Wasson for the temperature at a pressure of 10$^{-4}$ bar
at which 50 \% of the element is in solid compounds.
Of course, QY Sge's present atmosphere has a decidely non-solar
composition but a key influence on condensation of solids is where the C/O
ratio is placed with respect to the critical value C/O =1. QY Sge
has and presumably has had a C/O ratio of less than one, and, therefore,
the assumption of a solar mix is likely to be a valid approximation.
 Figure 7
shows QY Sge's [X/H] from Table 1 plotted against T$_{\rm cond}$.
There is a clear tendency for the elements of highest T$_{\rm cond}$
to be underabundant relative to those of low T$_{\rm cond}$. Carbon
and nitrogen should be discounted because they may
have been altered in the course of the star's evolution. Oxygen and sodium  may
also have been enriched. The initial metallicity of the star as indicated by
sulphur and zinc is close to  solar.
 It is notable that, in this
case and unlike a majority of the RV B variables, iron and other elements
are only slightly, if at all, depleted. The severe depletions are restricted to
Sc, Ti, and three heavy elements.
 The scatter at a fixed T$_{\rm cond}$ is
similar to that for  well observed RV Tauri RV B variables. 
There is a reasonable correlation between the abundances [X/H] and
the depletions measured for the cold interstellar diffuse clouds projected
in front of $\zeta$ Oph. This is shown in Figure 8 where the data for
$\zeta$ Oph are taken from Savage \& Sembach (1996). As noted above, nitrogen
should be discounted because it may be enriched in QY Sge. Scandium and
the heavy elements are not
plotted for lack of  estimates for $\zeta$ Oph. Aluminium probably
falls on the general trend: our estimate is consistent with 
Morton's (1985) result for $\zeta$ Oph of [X/H] of -3.1 to -3.4. 
For QY Sge, the heavy elements Y to Eu fit the
trend with T$_{\rm cond}$ quite well: Ba and Eu with T$_{\rm cond} \simeq
 1200$ K have approximately solar abundances but Y, Zr, and Ce with higher
T$_{\rm cond}$ are underabundant. 
There is no evidence from Figure 7
that QY Sge is enriched in the $s$-process elements but this too is
a common property of the RV Tauri variables.

\begin{figure}
\epsfxsize=8truecm
\epsffile{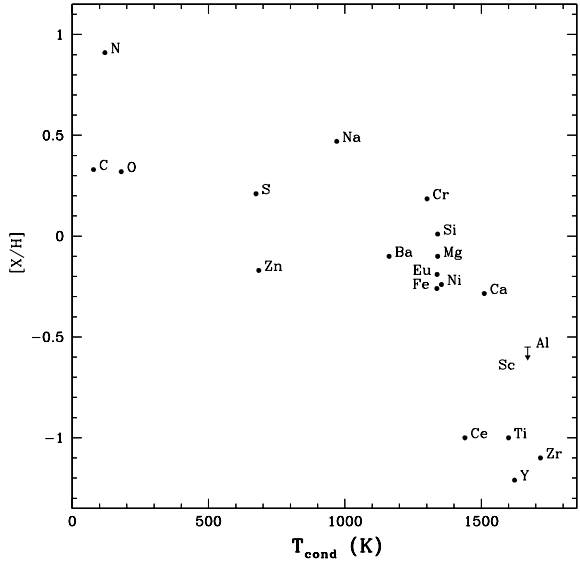}
\caption{The abundance [X/H] versus the condensation temperature T$_{\rm cond}$.
Elements are identified by their chemical symbols.}
\end{figure}

\begin{figure}
\epsfxsize=8truecm
\epsffile{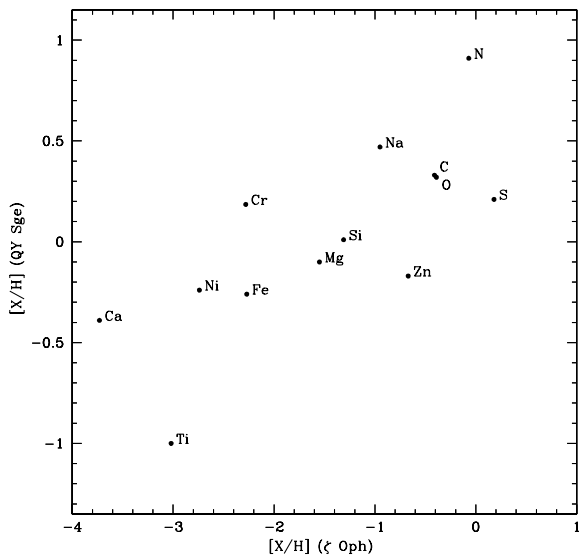}
\caption{The abundance [X/H] for QY Sge versus [X/H] for the cold interstellar
diffuse gas in front of $\zeta$ Oph.
Elements are identified by their chemical symbols.}
\end{figure}

Given the abundance anomalies, we
suppose that, as in the warmer RV Tauri variables,  dust
condenses in QY Sge's environs and is expelled by radiation pressure.
Not all the remaining gas is dragged away with the grains. Some of the gas is
able to return to the star. If this winnowing of dust from gas is
maintained for a long period, and, if mixing between the photosphere and
the envelope is suppressed, the photosphere assumes the composition of the
returning gas. QY Sge which certainly now maintains a nearby reservoir of
dust appears to have a photosphere reduced in condensible elements.
At the same time, the photosphere has approximately normal
abundances of elements like Si, Mg, and Fe which should be
principal constituents of  the circumstellar dust.
From Figure 7 we infer a dust temperature of about 1400 K for the regions of
effective winnowing. This temperature is considerably hotter than the
600 K inferred from the infrared excesss so that the effective
regions are probably interior to the exterior of the thick dusty layers
obscuring our direct view of the star.

\section{The Circumstellar Environment}

\begin{figure}
\epsfxsize=8truecm
\epsffile{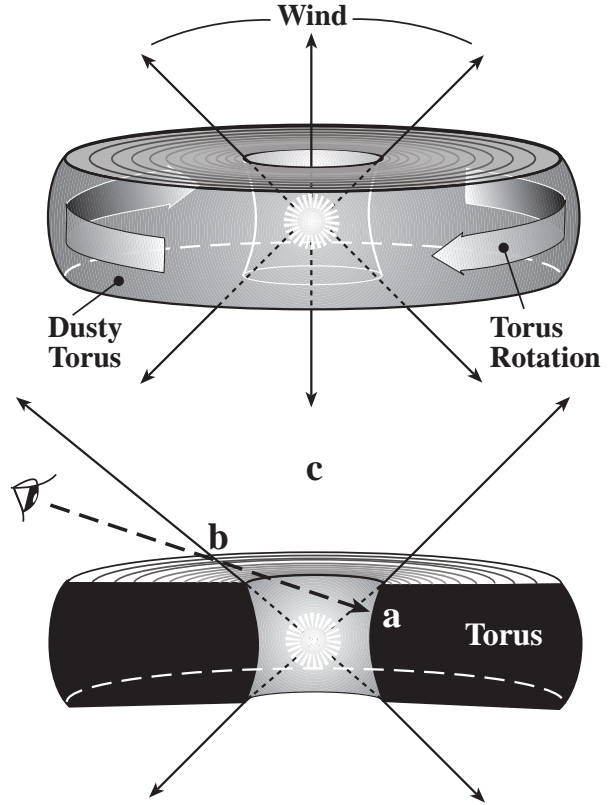}
\caption{Sketch of a possible geometry for QY Sge and its circumstellar
environment. Starlight reaches the observer after scattering off the
inner edge (region $a$)
 of the dusty rotating  torus. A bipolar wind off the star is the
site for the formation of the sharp emission lines near the star
(for example, the path length $a$ to  $b$)
 and broad emission lines farther from the star (region $c$).}
\end{figure}

Our observations provide several novel results to complement those presented
by Menzies \& Whitelock (1988) from low resolution spectra and photometry.
Their data led them to propose
 a model of `an extremely non-uniform circumstellar
shell' heated by starlight. In extending this idea, we must account for 
the characteristics of  the three distinct spectroscopic components:
\begin{itemize}
\item
The unusually broad photospheric
absorption lines. The breadth is presumed to arise from Doppler
shifts imposed by scattering off moving dust grains. 
\item
The sharp emission lines and
 the 
 absorption lines of
 the pumping resonance lines. The fact that absorption and emission
lines occur at the same radial velocity would suggest that the
emitting and absorbing volumes are coincident and along the
line of sight from the scattering surface to the observer. 
\item
The broad emission in the resonance lines of Na D,
K\,{\sc i}, Ca\,{\sc i} and {\sc ii}. 
 Lack of polarization of the
broad Na\,D emission lines implies 
that the emitted photons are not scattered from 
the surface contributing the photospheric spectrum. 
 Trammell
et al. (1994) in their polarization measurements
 did not resolve the Na\,D emission into its broad and sharp
components, but we assume that because the lion's share of the equivalent
width was most probably contributed by the broad component that 
it is unpolarized; the sharp component may also be unpolarized.
 We assume too that the
weaker broad lines are also umpolarized.
\end{itemize}

Adapting a geometry now recognized as not uncommon for the outer regions of
post-AGB stars and planetary nebulae, we place the bulk of the 
dust in an equatorial torus with its inner edge close to the star, a
requirement set by the derived black body dust temperature.
Along the line of sight to the star, the dust is
so very optically thick that directly transmitted photospheric
 light contributes   a negligible
fraction to the  observed optical spectrum.
If the line of sight is inclined slightly to the plane of the torus, the
rear of the torus' inner edge   will be visible but not the star (Figure 9).
In our picture, there is bipolar flow of gas
 from the star. It seems likely that the present wind from the 6000 K
star contains little or no dust. The dusty torus may be leftover
from the star's earlier life as a mass-losing AGB star. 
There is evidence that even more evolved (i.e., hotter)
 post-AGB stars with strong
infrared excesses are members of binary systems, e.g.,  89 Her (Walters
et al. 1993), and HR\,4049 (Bakker et al. 1998).
The geometry recalls that proposed for VY CMa, a mass-losing M supergiant,
by Herbig (1969, 1970).

Qualitatively, the origins of the three distinct spectroscopic features may be
placed as follows:
\begin{itemize}
\item
The rear of the inner edge of the dusty torus
is assumed primarily responsible for scattering the photospheric spectrum toward the
observer.
On the assumption that the dust is approximately in Keplerian motion, the line
broadening will result from the Doppler shifts imposed by scattering over an extended
part of the inner edge. The net Doppler shift should be close to zero.
\item
Gas near the star between the observer and the torus' inner edge is exposed to
starlight. This gas should be at a low density such that
 most atoms and ions will be in their  ground states.
These atoms may be excited to give the sharp emission lines by photons
received directly from the star or indirectly by photons scattered off the
torus. Presumably the former are more numerous. In this case, the
profile of the sharp emission line is set by the wind  velocity (assumed radial),
the opening angle of the bipolar flow, and the angle of inclination of the torus
to the line of sight. A  resonance line seen in absorption in the spectrum of the
scattered light from the torus' inner edge will  show the same spread in 
Doppler shifts as a sharp emission line. To within the measurement errors, the
sharp emission and their exciting absorption lines do have the same width.
Additional absorption in resonance lines may occur in the path between the star
and the point of scattering off the torus. This absorption component will be
Doppler-shifted to the blue relative to the scattered stellar
light, and the  accompanying emission, if visible to the
observer, will be red-shifted. 
\item
Broad resonance emission lines are assigned to resonance fluorescence by
stellar photons interacting with atoms in  the  bipolar flow. If the torus is
of large radius, the emission from the receding half of the bipolar flow will
be occulted, and the observed emission line will show a small net blue-shift with
a magnitude dependent on the wind velocity, and
the tilt between the axis of the bipolar flow and the
line of sight. According to Figure 9, an emission line's width
 will depend on the same factors; 
emitting gas far from the torus will have line of sight velocities from
approximately $-v^g_w\cos\theta$ to $+v^g_w\cos\theta$,
 where $v^g_w$ is the velocity of the
gas at that radial distance from the star, and
$2\theta$ is the angle subtended at the star by the inner edge of the
torus. The observed line width
implies a wind velocity of about 100 - 150 km s$^{-1}$.
Since the torus is here considered to be the polarizing source for the
scattered photospheric spectrum, the Na\,D and other broad emission lines will
be polarized differently and probably to a significantly smaller degree.
\end{itemize}

The model makes qualitative predictions about velocity shifts relative to
the systemic stellar velocity. At this time, too few observations of the
stellar absorption lines are available to determine the systemic
velocity. Measured velocities are also affected by possible changes in
the net velocity imposed by the scattering surface. It should be noted that
the model does not explain readily why the emission lines are either sharp or
broad. This is particularly an issue for the broad lines which are
obviously accompanied by a sharp line (Figure 3). Our model supposes
both lines are formed in the bipolar flow. It seems necessary to invoke
a sharp transition from a slow to a high velocity wind but this should
occur without a large change in physical conditions of the gas;
 the sharp lines are
formed below the transition point and the broad  lines above.

\section{Concluding Remarks}

QY Sge is here identified as descended from a luminous red giant
on either the higher reaches of the red giant branch (RGB) prior to
helium ignition or the AGB.
Assuming a mass of about 1$M_\odot$, QY Sge's T$_{\rm eff}$ and $\log g$
correspond to a luminosity $L \simeq 10^4L_\odot$. This is achieved
by a star near the tip of the RGB or
 at a quite advanced along the AGB (Figure 6).
Carbon enrichment of the surface following the expected depletion
resulting from the first dredge-up early on the RGB could possibly
have occurred during the He-core flash at the tip of the RGB but more
likely is a result of the third dredge-up early on the AGB; the 
star evolved away from the AGB
 before
the $s$-process manufactured heavy elements in large quantities
in the He-shell.

QY Sge is likely quite distant from the Sun.
Menzies \& Whitelock used the apparent bolometric magnitude  from the
flux curve in combination with assumed absolute bolometric
magnitudes to arrive at a distance. 
For example, adopting their $m_{\rm bol} \sim 8.5$ and 
including a small bolometric correction, our luminosity  estimate puts the
distance to QY Sge at about 3 kpc. This is an upper limit because
the dust cannot block all of the stellar radiation, and, furthermore,
the optical fluxes depend on the efficiency of the scattering of
light by the less dusty regions. 
Interstellar absorption is small. Menzies \& Whitelock estimated
$A_V \sim 1.2$ mag from the B - V and the intrinsic colours of a G0I
star, and noted that reddening maps suggest $A_V \sim 2$ for distances
greater than about 600 pc. An independent estimate of reddening is
possible from our detection of the diffuse interstellar band at
6284 \AA. The band's equivalent width of 461 m\AA\ with a calibration
offered by Somerville (1995) gives $E(B - V) \simeq$ 0.40, or $A_V \simeq
1.2$. 
Circumstellar absorption for the direct light from the star is
substantial; we estimate about 6 magnitudes for the visual from comparison
of the infrared (I,J,H) colours with the colours of unreddened
G supergiants.

QY Sge, as noted above, shares T$_{\rm eff}$, $\log g$, and abundance
anomalies with the RV Tauri variables. This association suggests that
stars may commonly depart the AGB due to severe mass loss and that
QY Sge is not the result of freak circumstances.
A link with the RV\,Tauri variables recalls the  suggestion
that all RV Tauri variables belong to binary systems with 
substantial circumbinary dusty disks
(Van Winckel et al. 1999).
In addition, 
QY Sge has many similarities with 89 Her, a spectroscopic binary with
a very low mass companion to the post-AGB primary: similar H$\alpha$ profiles, similar
sharp low excitation emission lines of neutral metals, and a considerable
infrared excess (Walters et al. 1993). There is one prominent difference between the pair:
89 Her shows a blue-shifted high velocity absorption at Na D in
place of QY Sge's broad emission lines, but,  according to our
model, QY Sge would show such absorption if seen pole-on.

 Further photometric and spectroscopic
observations are desirable to investigate  whether QY Sge is a
RV Tauri variable and/or a spectroscopic binary. At present, we cannot guess the systemic velocity. 
Higher quality spectra would be useful in order to effect line profile
comparisons of emission and absorption lines for the low excitation
transitions. Inflow of gas onto the star may be occurring in order to
sustain the abundance anomalies (Figure 7). Is gas from the inner edge of the
torus falling onto the star? If so, high-resolution high S/N spectra around
key resonance lines may reveal the infalling gas in the spectrum of the
light scattered off the torus' inner edge. This would be the first detection
of such a gas stream in stars affected bya winnowing of gas from dust.

Our attention was drawn to QY Sge by the report of broad 
Na D emission lines. Very few stars show such lines in their spectra.
Apart from the R CrB stars observed in deep minimum, QY Sge may have
very few cousins;  we noted VY CMa and
Lloyd  Evans (1997) mentions IRAS 13258-8103 as
having strong Na D emission. The geometry of the
circumstellar material suggested for QY Sge may be a common one.
Menzies \& Whitelock point out that, if viewed from another angle, QY Sge
would be observed as a normal supergiant with a relatively
weak infrared excess. `Perspective effects' give QY Sge its
prominent infrared excess and strong broad emission lines.
This geometry would in principle account for the lines in
the R CrB stars but they also show high-excitation He\,{\sc i}
(and other) lines.

\section{Acknowledgements}
We thank Gajendra Pandey, Eswar Reddy, and David Yong for observing QY Sge at our
request,  Sunetra Giridhar for providing Figure 6, and Melody Lambert
for designing Figure 9.
This research has been supported in part by the US National Science
Foundation (grant AST 9618414).

\end{document}